\documentclass[a4paper,10pt]{article}%
\usepackage{geometry}
\usepackage{amsmath}
\usepackage{amsfonts,pslatex}
\usepackage{amssymb}
\usepackage{graphicx}%
\usepackage[small,bf]{caption}
\begin{document}

\date{}
\title{\textbf{The (IR-)relevance of the Gribov ambiguity in $SU(2)\times U(1)$ gauge theories with fundamental Higgs matter}}
\author{\textbf{M.~A.~L.~Capri}$^{a}$\thanks{caprimarcio@gmail.com}\,\,,
\textbf{D.~Dudal}$^{b}$\thanks{david.dudal@ugent.be}\,\,,
\textbf{M.~S.~Guimaraes}$^{a}$\thanks{msguimaraes@uerj.br}\,\,,
\textbf{I.~F.~Justo }$^{a}$\thanks{igorfjusto@gmail.com}\,\,,
\textbf{S.~P.~Sorella}$^{a}$\thanks{sorella@uerj.br}\,\,,
\textbf{D.~Vercauteren}$^{a}$\thanks{vercauteren.uerj@gmail.com}\, \thanks{Work supported by
FAPERJ, Funda{\c{c}}{\~{a}}o de Amparo {\`{a}} Pesquisa do Estado do Rio de
Janeiro, under the program \textit{Cientista do Nosso Estado}, E-26/101.578/2010.}\,\,\\[2mm]
{\small \textnormal{$^{a}$  \it Departamento de F\'{\i }sica Te\'{o}rica, Instituto de F\'{\i }sica, UERJ - Universidade do Estado do Rio de Janeiro,}}
 \\ \small \textnormal{\phantom{$^{a}$} \it Rua S\~{a}o Francisco Xavier 524, 20550-013 Maracan\~{a}, Rio de Janeiro, Brasil}\\
	 \small \textnormal{$^{b}$ \it Ghent University, Department of Physics and Astronomy, Krijgslaan 281-S9, 9000 Gent, Belgium}\normalsize}
\maketitle

\begin{abstract}
It is well accepted that dealing with the Gribov ambiguity has a major impact on correlation functions in gauge-fixed Yang--Mills theories, in particular in the low momentum regime where standard perturbation theory based on the Faddeev--Popov approach fails. Recent results, derived from functional tools (Dyson--Schwinger equations or exact RG) or the effective Gribov--Zwanziger action method, pointed towards e.g. gauge boson correlation functions that are not compatible with the properties of observable degrees of freedom. Although such an observation is a welcome feature for gauge theories exhibiting confinement, it would be a discomfort for gauge theories supplemented with Higgs fields, cfr.~the experimental success of the electroweak model based on a $SU(2)\times U(1)$ gauge group. The purpose of this short note is to assure that the effective action resolution to the Gribov ambiguity reduces to the standard Faddeev--Popov method in the \emph{perturbative} regime of sufficiently small coupling/large Higgs condensate, thereby not compromising the physical particle spectrum of massive gauge bosons and a massless photon for the $SU(2)\times U(1)$ gauge-Higgs model. The closer the theory gets to the limit of vanishing Higgs condensate, the more the Gribov problem resurfaces with all its consequences. We give some speculations w.r.t.~the Fradkin--Shenker insights about the phase diagram.
\end{abstract}


\section{Motivation}
The continuum path integral or canonical Hamiltonian quantization of gauge theories  require the choice of a gauge fixing condition. This is perfectly well understood in the Abelian case, but in the non-Abelian case the nontrivial topology of the gauge group can spoil the simple picking up of a unique gauge field per set of gauge equivalent fields that fulfills the gauge condition. This problem was first investigated by Gribov in \cite{Gribov:1977wm} for the Landau and Coulomb gauges, and later on many nontrivial results were more rigorously probed in the work of Zwanziger, Dell'Antonio and others \cite{Zwanziger:1989mf}. Also in curved spaces, some results were booked in more recent times \cite{Canfora:2011xd}.

In a series of papers, we and others investigated the dynamical stability and further quantum alterations to the seminal results of Gribov and Zwanziger \cite{Dudal:2008sp}.
The ensuing propagators of the gauge bosons (gluons) and Faddeev-Popov ghosts are in good agreement with their lattice counterparts \cite{Cucchieri:2011ig}. The Gribov--Zwanziger approach is thus one example of effective functional approaches to non-Abelian gauge fixed theories, thereby supplementing other potential approaches as in \cite{Aguilar:2013vaa}. The propagators proposed by us recently also found use in studies of Casimir effect \cite{Canfora:2013zna} or the finite temperature phase diagram \cite{Canfora:2013kma,Fukushima:2012qa}.

A key observation of the nonperturbative Landau gauge gluon propagator is the violation of positivity, that is the gluon is not a particle that has a physical interpretation with associated K\"{a}ll\'{e}n-Lehmann spectral integral representation with positive density. This violation of positivity is well-appreciated from simulational and analytical approaches \cite{Bowman:2007du}, and it can be seen as a signal of confinement: gluons cannot be observed. In the Gribov--Zwanziger approach, this violation of positivity annex unphysicalness of the gluon is imminent because of the presence of complex conjugate poles in the propagator \cite{Baulieu:2009ha}.

This led us to the interesting question what the consequences would be of the gauge fixing ambiguity and its Gribov-Zwanziger resolution when the gauge theory is coupled to a Higgs sector, most notably in the case of a $SU(2)\times U(1)$ theory with fundamental Higgs matter, of physical relevance to the electroweak sector of the Standard model. At the energies at which the electroweak theory is probed, a valid perturbation theory with a set of 3 observable massive gauge bosons and a massless photon should be the outcome. Complex conjugate masses or a positivity violation in the gauge boson sector should thus preferably not manifest themselves. Of course, the Gribov problem is still there, irrespective whether Higgs field are present or whether the Higgs mechanism takes place. The issue of having multiple solutions to the gauge fixing condition is a classical observation, only influenced by having a non-Abelian gauge field structure. We would thus rather like to find out under which circumstances treating the Gribov copy problem becomes trivial at the quantum level, thereby not affecting the standard perturbative expansion and accompanying results on the spectrum etc.

The difference or not between a confining and Higgs-like spectrum of gauge-Higgs system is a very deep question, most notable brought to attention in the seminal theoretical lattice work of Fradkin and Shenker \cite{Fradkin:1978dv}. They presented evidence, using either an area or perimeter law for the Wilson loop expectation value to label the confining and Higgs phase, that both regions are analytically connected in a (gauge coupling, Higgs $vev$) diagram. From lattice works as \cite{Caudy:2007sf}, one also learns that the 2 regions are separated by a line corresponding to a 1st order phase transition, whereby it is crucial to mention that this line has an end-point. The spectrum is in both phases generated from gauge invariant operators that ``interpolate'' analytically between massive gauge bosons, resp.~mesonlike bound states in the Higgs, resp.~confining phase, see for example also \cite{frohlich}. We refer to \cite{Fradkin:1978dv,Greensite:2011zz} for a more detailed exposure on this.

\section{Setup} \label{sect2}
In this section we define the notation and conventions used throughout the paper. The system we study is the electroweak sector of the $4d$ standard model, thereby generalizing our earlier works \cite{Capri:2012cr,Capri:2012ah,threedcase}. In the following sections we consider in detail the possible effects of Gribov copies in such a setting, based on the original Gribov no-pole analysis \cite{Gribov:1977wm,Sobreiro:2005ec}. A summary of our results can be found in section 6.
\subsection{The classical action}
Let us begin by recalling the action describing a Higgs field in the fundamental representation of $SU(2)$:
\begin{equation}
\label{Sf}
S=\int d^{4}x\left(\frac{1}{4}F_{\mu \nu }^{a}F_{\mu \nu }^{a}+ \frac{1}{4} B_{\mu\nu} B_{\mu\nu} +
(D_{\mu }^{ij}\Phi ^{j})^{\dagger}( D_{\mu }^{ik}\Phi ^{k})+\frac{\lambda }{2}\left(
\Phi ^{\dagger}\Phi-\nu ^{2}\right) ^{2}\right)  \;,
\end{equation}
with covariant derivative
\begin{equation}
D_{\mu }^{ij}\Phi^{j} =\partial _{\mu }\Phi^{i} - \frac{ig'}{2}B_{\mu}\Phi^{i} -   ig \frac{(\tau^a)^{ij}}{2}A_{\mu }^{a}\Phi^{j}
\end{equation}
and the vacuum expectation value (\textit{vev}) of the Higgs field
\begin{equation}
\langle \Phi \rangle  = \left( \begin{array}{ccc}
                                          0  \\
                                          \nu
                                          \end{array} \right)  \;.
\label{vevf}
\end{equation}
The indices $i,j=1,2$ refer to the fundamental representation of $SU(2)$, and $\tau^a$, $a=1,2,3$, are the Pauli matrices. The coupling constants $g$ and $g'$ refer to the groups $SU(2)$ and $U(1)$ respectively. It will be understood that we work in the limit $\lambda\to\infty$ for simplicity, i.e.~we have a Higgs field frozen at its vacuum expectation value, as in \cite{Fradkin:1978dv} and as in our previous works. In principle, in the continuum, such limit should only be taken after the Feynman diagrams up to a certain order have been computed and properly renormalized, since otherwise problems with the UV renormalization can be expected. In this work, as in its predecessors \cite{Capri:2012cr,Capri:2012ah,threedcase}, we shall however only need a one loop computation for the ghost propagator (see further) and at this order, there are no contributions proportional to $\lambda$ yet, so the limit $\lambda\to\infty$ is well under control for the time being. In any case, the aim of the current paper is still not that of being exhaustive on the Gribov/Higgs issue, given the mere complexity of it. The to be presented results should be interpreted as a first step towards a better understanding of the gauge fixing ambiguity in gauge theories supplemented with Higgs matter fields.

The respective field strengths are defined as
\begin{equation}
F^a_{\mu\nu} = \partial_\mu A^a_\nu -\partial_\nu A^a_\mu + g \varepsilon^{abc} A^b_\mu A^c_\nu \;, \qquad B_{\mu \nu} = \partial_\mu B_\nu -\partial_\nu B_\mu  \;.
\label{fs}
\end{equation}
For later usage, consider quadratic part of the action \eqref{Sf}, which reads
\begin{eqnarray}
S_\text{quad} &=& \int\!\! d^{4}x \, \frac{1}{2}A_{\mu}^{\alpha}\left[ \left(-\partial_{\mu}\partial_{\mu} + \frac{\nu^{2}}{2}g^{2}\right)\delta_{\mu\nu} + \partial_{\mu}\partial_{\nu} \right]A_{\nu}^{\alpha} + \int\!\! d^{4}x\, \frac{1}{2}B_{\mu}\left[ \left(-\partial_{\mu}\partial_{\mu} + \frac{\nu^{2}}{2}g'^{2}\right)\delta_{\mu\nu} + \partial_{\mu}\partial_{\nu}\right]B_{\nu}  \nonumber \\
&+& \int\!\! d^{4}x\, \frac{1}{2}A_{\mu}^{3}\left[ \left(-\partial_{\mu}\partial_{\mu} + \frac{\nu^{2}}{2}g^{2}\right)\delta_{\mu\nu} + \partial_{\mu}\partial_{\nu}\right]A_{\nu}^{3} - \frac{1}{4}\int\!\! d^{4}x\, \nu^{2}g\,g'\,A^{3}_{\mu}B_{\nu} - \frac{1}{4}\int\!\! d^{4}x\, \nu^{2}g\,g'\,B_{\mu}A^{3}_{\nu} \;.
\label{Squad}
\end{eqnarray}
Making use of
\begin{equation} \label{wsandza}
W^+_\mu = \frac{1}{\sqrt{2}} \left( A^1_\mu + iA^2_\mu \right) \;, \quad W^-_\mu = \frac{1}{\sqrt{2}} \left( A^1_\mu - iA^2_\mu \right)  \;, \quad
Z_\mu =\frac{1}{\sqrt{g^2+g'^2} } \left(  -g A^3_\mu + g' B_\mu \right) \;, \quad A_\mu =\frac{1}{\sqrt{g^2+g'^2} } \left(  g' A^3_\mu + gB_\mu \right) \;,
\end{equation}
which has the inverse transformation
\begin{equation}
A^1_\mu = \frac{1}{\sqrt{2}} \left( W^+_\mu + W^-_\mu \right) \;, \quad A^2_\mu = \frac{1}{i\sqrt{2}} \left( W^+_\mu - W^-_\mu \right) \;, \quad
B_\mu =\frac{1}{\sqrt{g^2+g'^2} } \left(  g A_\mu + g' Z_\mu \right) \;, \quad A^3_\mu =\frac{1}{\sqrt{g^2+g'^2} } \left(  g' A_\mu - gZ_\mu \right) \;.
\end{equation}
one easily derives that $S_\text{quad}$ is diagonal:
\begin{eqnarray}
S_\text{quad} & = & \int d^4 x   \left( \frac{1}{2} (\partial_\mu W^+_\nu - \partial_\nu W^+_\mu)(\partial_\mu W^-_\nu - \partial_\nu W^-_\mu)  + \frac{g^2\nu^2}{2}W^+_\mu W^-_\mu   \right)  \nonumber \\
&+ &\int d^4x  \left(  \frac{1}{4} (\partial_\mu Z_\nu - \partial_\nu Z_\mu)^2  + \frac{(g^2+g'^2)\nu^2}{4}Z_\mu Z _\mu  +    \frac{1}{4} (\partial_\mu A_\nu - \partial_\nu A_\mu)^2  \right)  \;.
\label{qd}
\end{eqnarray}

\subsection{Gauge fixing, the Gribov ambiguity and the restriction to the Gribov region}
\label{TRGR}
As we are working in a continuum setting, we need to restrict the local gauge freedom using a suitable gauge fixing. Here, we adopt the Landau gauge, i.e.~we demand that all gauge fields be transverse (= vanishing divergence). The gauge fixing action can be written as
\begin{equation}
S_\text{gf} = \int d^4x \; \left( b^a \partial_\mu A^a_\mu + {\bar c}^a \partial_\mu D^{ab}_\mu c^b
+ b\partial_\mu B_\mu + {\bar c}\partial^2 c \right) \;,
\label{cgf}
\end{equation}
after a suitable linear redefinition of the necessary Lagrange multipliers and anti-ghosts.  The corresponding gauge field propagators read
\begin{subequations} \label{AB gluon prop} \begin{gather}
\langle A^{\alpha}_{\mu}(p) A^{\beta}_{\nu}(-p) \rangle = \frac{1}{p^{2} + \frac{\nu^{2}}{2}g^{2}}T_{\mu\nu}(p^2)\delta^{\alpha \beta} \;, \qquad
\langle A^{3}_{\mu}(p) A^{3}_{\nu}(-p) \rangle = \frac{1}{p^{2} + \frac{\nu^{2}}{2}(g^{2} + g'^{2})} T_{\mu\nu}(p^2)\;,  \\
\langle B_{\mu}(p) B_{\nu}(-p) \rangle = \frac{1}{p^{2}} T_{\mu\nu}(p^2) \;,  \qquad
\langle A^{3}_{\mu}(p) B_{\nu}(-p) \rangle = \frac{\nu^{2}}{4}(g^{2} + g'^{2}) \frac{g'^{2}}{g\,g'} \frac{1}{p^{2}\left(p^{2} + \frac{\nu^{2}}{2}(g^{2} + g'^{2}) \right)} T_{\mu\nu}(p^2)\;.
\end{gather} \end{subequations}
The indices $\alpha$ and $\beta$ take the values 1 and 2 and denote the off-diagonal modes. We also introduced the transversal projector $T_{\mu\nu}(p^2)=\left(\delta_{\mu\nu}-\frac{p_\mu p_\nu}{p^2}\right)$. Rewriting these propagators in terms of the physical fields $W^{\pm}_{\mu}$, $Z_{\mu}$ and $A_{\mu}$ gives
\begin{equation}
\label{WZAgluonprop1} \langle W^{+}_{\mu}W^{-}_{\nu} \rangle = \frac{1}{p^{2} + \frac{\nu^{2}}{2}g^{2}}T_{\mu\nu}(p^2) \;, \qquad
\langle Z_{\mu}(p) Z_{\nu}(-p) \rangle = \frac{1}{p^{2} + \frac{\nu^{2}}{2}(g^{2} + g'^{2})} T_{\mu\nu}(p^2)\;, \qquad
\langle A_{\mu}(p) A_{\nu}(-p) \rangle = \frac{1}{p^{2}} T_{\mu\nu}(p^2) \;.
\end{equation}
Now, it is well known since the pioneering paper \cite{Gribov:1977wm} that imposing the Landau gauge fixing does not completely fix the gauge freedom. Indeed, since an infinitesimal gauge transformation corresponds to a shift of the gauge fields with the covariant derivative of some function, a solution gauge equivalent to the Landau gauge ---with infinitesimal gauge transformation--- is possible whenever the Faddeev--Popov operator has zero modes, that is if the Jacobian of the gauge fixing has zero modes. Since the Jacobian matrix enters the Faddeev--Popov quantization procedure explicitly, it is clear that the latter is incompatible with having gauge copies.

Gribov's original proposal was to restrict the Landau gauge configurations to the subset $\Omega$ defined by the Faddeev--Popov operator $-\partial_\mu D_\mu^{ab}$ having no zero modes. Needless to say, this at least kills off the infinitesimal gauge copies. This approach was later put on a firmer footing by Zwanziger, see \cite{Vandersickel:2012tz} for a recent review with a myriad of relevant references. Related or somewhat different approaches can be found in e.g.~\cite{Serreau:2012cg,Zwanziger:2001kw}.

We will implement the restriction to the Gribov region $\Omega$ in terms of the original fields $A^a_\mu$ and $B_\mu$ and only afterwards move to the other fields $W^+$, $W^-$, $Z$, and $A$. This simplifies the calculations of the Gribov form factors, see next section. From now on, we shall be working with expression \eqref{cgf}.
Adding the gauge fixing action to the expression (\ref{Sf}), the total action reads
\begin{equation}
S = \int d^{4}x \left[\frac{1}{4} F^{a}_{\mu \nu}F^{a}_{\mu \nu} + \frac{1}{4} B_{\mu \nu}B_{\mu \nu} + \bar{c}^{a}\partial_{\mu}D_{\mu}^{ab}c^{b} + \bar{c} \partial^{2}c + b^{a}\partial_{\mu}A^{a}_{\mu} + b\partial_{\mu}B_{\mu} + \left(D_{\mu}^{ij}\phi^{j}\right)^{\dagger} \left(D_{\mu}^{il}\phi^{l}\right) + \frac{\lambda}{2} \left( \phi^{\dagger}\phi - \nu^{2}\right)^{2}\right]. \label{Stotal}
\end{equation}
It is interesting to notice here that in general, thanks to the transversality of the Landau gauge, there will be no mixing terms between the massive gauge bosons and associated Goldstone mode fields, due to the presence, after partial integration of a $\partial_\mu A_\mu$ factor. This is not a surprise, since after all the Landau gauge is an extremal case of the well-known $R_\xi$ gauges which have the explicit property of killing the mixing terms between the gauge bosons and Goldstones. The diagrammatic expansion of the here studied $SU(2)\times U(1)$ theory should thus be well under control. We do however wish to remark that the issue of unitarity of gauge theories with massive vector particles due to a Higgs mechanism, is ideally not discussed in the Landau gauge, mainly because of the constraint issues related to setting $\partial_\mu A_\mu=0$ with the help of a Lagrangian multiplier $b$. Other gauges are better suited for this, in particular the well-known unitary gauge. On the other hand, this gauge might provide with nice physical interpretation at the classical level, it is evenly well-known that the UV renormalization of the unitary gauge is badly behaved, to say the least. In this work, we thus specifically choose the Landau gauge, since it is the only covariant gauge in which relatively a lot is known about the Gribov gauge fixing ambiguity, while being a well-defined gauge at the quantum level for carrying out loop computations. In addition, it also has the advantage of being under scrutiny using lattice simulational tools, see \cite{Maas:2010nc}, allowing a comparison between different methodologies. In theory, if unitarity is established in one gauge, gauge invariance dictates it is valid in any gauge. At the quantum level, the unitarity is always better discussed using BRST tools anyhow \cite{Becchi:1974xu,Dudal:2012sb}. We will not dwell upon that here.

To restrict to the first Gribov region $\Omega$ at lowest nonvanishing order, we can compute the two-point ghost function and impose a no-pole condition for the latter quantity. Indeed, the ghost propagator is nothing but the inverse of the Faddeev--Popov operator. The tree level ghost propagator reads $\frac{1}{k^2}$ and is thus positive. If the perturbative corrections get too large --- which we typically expect in strongly interacting gauge theories to happen at lower momentum transfer --- the quantum shift in the ghost self-energy could overcome the tree level value. In other words, the ghost propagator would turn negative, indicating that we are outside of the region $\Omega$, defined by the ghost propagator being positive.

\section{The no-pole condition made explicit}
In order to impose the no-pole condition on the SU(2) ghost propagator, we first need a closed expression for it. The two-point ghost correlation function is
\begin{equation}
\mathcal{G}^{ab}(k;A) = \frac{1}{k^{2}}\left[ \delta^{ab} - g^{2}\frac{k_{\mu}k_{\nu}}{k^{2}} \int\!\! \frac{d^{4}p}{(2\pi)^{4}} \,\, \epsilon^{amc}\epsilon^{cnb} \frac{1}{(k-p)^{2}}A^{m}_{\mu}(p)A^{n}_{\nu}(-p)\right]
\label{ghprop}
\end{equation}
which can be written in matrix-form as
\begin{equation}
\mathcal{G}^{ab}(k;A) = \left(
  \begin{array}{ll}
   \delta^{\alpha \beta} \mathcal{G}_\text{off}(k;A) & \,\,\,\,\,\,\,\,0 \\
   \,\,\,\;\;\;\;\;\;0 & \mathcal{G}_\text{diag}(k;A)
  \end{array}
\right),
\label{gh prop offdiag}
\end{equation}
Here we defined
the off-diagonal two-point function
\begin{subequations} \begin{equation}
\mathcal{G}_\text{off}(k;A) = \frac{1}{k^{2}} \left[1+ g^{2}\frac{k_{\mu} k_{\nu}}{k^{2}} \int\!\! \frac{d^{4}p}{(2\pi)^{4}} \frac{1}{(k-p)^{2}} \frac{1}{2} \left(A^{\alpha}_{\mu}(p)A^{\alpha}_{\nu}(-p) + 2A^{3}_{\mu}(p)A^{3}_{\nu}(-p)\right)\right]
\label{gh off}
\end{equation}
and the diagonal one
\begin{equation}
\mathcal{G}_\text{diag}(k;A) = \frac{1}{k^{2}} \left[1+ g^{2}\frac{k_{\mu} k_{\nu}}{k^{2}} \int\!\! \frac{d^{4}p}{(2\pi)^{4}} \frac{1}{(k-p)^{2}} A^{\alpha}_{\mu}(p)A^{\alpha}_{\nu}(-p)\right]\;.
\label{gh diag}
\end{equation} \end{subequations}
Thus
\begin{equation}
\mathcal{G}_\text{off}(k;A) \simeq \frac{1}{k^{2}} \left( \frac{1}{1 - \sigma_\text{off}(k;A)} \right) \;, \qquad \mathcal{G}_\text{diag}(k;A) \simeq \frac{1}{k^{2}} \left( \frac{1}{1 - \sigma_\text{diag}(k;A)} \right)\;.
\label{gh nopole}
\end{equation}
The quantities $\sigma_\text{off}(k;A)$ and $\sigma_\text{diag}(k;A)$ turn out to be decreasing functions of $k$, explicitly visible upon taking the average with a suitable measure, see also \cite{Gribov:1977wm}. Thus, the no-pole condition is eventually satisfied by the conditions
\begin{equation} \label{sigmanopoles}
\sigma_\text{off}(0;A) < 1 \;, \qquad \sigma_\text{diag}(0;A) < 1\;,
\end{equation}
where
\begin{subequations} \label{sigmaoffanddiag} \begin{equation}
\sigma_\text{off}(0;A) = \frac{g^{2}}{4} \int\!\! \frac{d^{4}p}{(2\pi)^{4}} \frac{1}{p^{2}} \left( \frac{1}{2} A^{\alpha}_{\mu}(p)A^{\alpha}_{\mu}(-p) + A^{3}_{\mu}(p)A^{3}_{\mu}(-p)\right)
\label{sigma off}
\end{equation}
and
\begin{equation}
\sigma_\text{diag}(0;A) = \frac{g^{2}}{4} \int\!\! \frac{d^{4}p}{(2\pi)^{4}} \frac{1}{p^{2}} A^{\alpha}_{\mu}(p)A^{\alpha}_{\mu}(-p)\;.
\label{sigma diag}
\end{equation} \end{subequations}
Conditions \eqref{sigmanopoles} imply that the ghost propagators $\mathcal{G}_\text{off}(k;A)$ and   $\mathcal{G}_\text{diag}(k;A)$ are always positive, thus ensuring that one remains inside the Gribov region $\Omega$. We employed the property
\begin{equation}
A_{\mu }^{a}(q)A_{\nu }^{a}(-q) =T_{\mu\nu}(q^2) \omega (A)(q)  \quad \Rightarrow \omega (A)(q)=\frac{1}{3}A_{\lambda }^{a}(q)A_{\lambda }^{a}(-q)  \label{p1}
\end{equation}
which follows from the transversality of the gauge field, $q_\mu A^a_\mu(q)=0$. Using rotational invariance, for a generic function $\mathcal{F}(p^2)$, we may always use
\begin{equation}
\int \frac{d^{4}p}{(2\pi )^{4}}T_{\mu\nu}(p^2) \mathcal{F}(p^2)=\mathcal{A}\;\delta _{\mu \nu }  \label{p2}
\end{equation}%
where, upon contracting both sides of equation \eqref{p2} with $\delta_{\mu\nu}$,
\begin{equation}
\mathcal{A}=\frac{3}{4}\int \frac{d^{4}p}{(2\pi )^{4}}\mathcal{F}(p^2)\;.   \label{p3}
\end{equation}

\subsection{Lowest-order gap equations from the no-pole condition}

In order to restrict to the Gribov region $\Omega$, we implement the no-pole conditions \eqref{sigmanopoles} by means of suitable step functions in the functional integral, namely  \begin{equation}
Z_\text{quad} = \mathcal{N} \int \mathcal{D}A\, \mathcal{D}B \,\theta(1-\sigma_\text{off})\, \theta(1-\sigma_\text{diag})\, e^{-S_\text{quad}},
\label{Zquad}
\end{equation}
where $S_\text{quad}$ is the quadratic part of the total action (\ref{Stotal}). Making use of the integral representation
\begin{equation}
\theta(x) = \int_{-i\infty +\epsilon}^{i\infty +\epsilon} \frac{d\omega}{2\pi i \omega} \; e^{\omega x}   \;, \label{rp}
\end{equation}
we get for the functional integral
\begin{eqnarray}
Z_\text{quad} &=& \mathcal{N} \int \frac{d \omega}{2\pi i \omega}\frac{d \beta}{2\pi i \beta} \mathcal{D}A\, \mathcal{D}B e^{\omega (1-\sigma_\text{off})} \, e^{\beta (1-\sigma_\text{diag})} e^{-S_\text{quad}} \nonumber \\
&=& \mathcal{N} \int \frac{d \omega}{2\pi i \omega}\frac{d \beta}{2\pi i \beta} \mathcal{D}A^{\alpha} \mathcal{D}A^{3} \, \mathcal{D}B\; e^{\omega}\,e^{\beta} \exp \left\{-\frac{1}{2} \int\!\! \frac{d^{4}p}{(2\pi)^{4}}\, A^{\alpha}_{\mu}(p) \left[ \left(p^{2} + \frac{\nu^{2}}{2}g^{2} + \frac{g^{2}}{2}\left(\beta +\frac{\omega}{2}\right)\frac{1}{p^{2}}\right) T_{\mu\nu}(p^2)\right]A^{\alpha}_{\nu}(-p) \right. \nonumber \\ && + A^{3}_{\mu}(p) \left[ \left(p^{2} + \frac{\nu^{2}}{2}g^{2} + \frac{g^{2}}{2}\frac{\omega}{p^{2}}\right) \times T_{\mu\nu}(p^2) \right]A^{3}_{\nu}(-p)  + B_{\mu}(p) \left[ \left(p^{2} + \frac{\nu^{2}}{2}g'^{2}\right)\left(\delta_{\mu\nu} - \frac{p_{\mu}p_{\nu}}{p^{2}}\right) \right] B_{\nu}(-p) \nonumber \\
&{}& \left. - A^{3}_{\mu}(p) \left[\nu^{2}g\,g'T_{\mu\nu}(p^2) \right]B_{\nu}(-p) \right\}\;.
\label{Zquad1}
\end{eqnarray}
It turns out to be convenient to perform the change of variables $\beta \to \beta -\frac{\omega}{2}$, $\omega \to \omega$, leading us to
\begin{eqnarray}
Z_\text{quad} &=&  \mathcal{N} \int \frac{d \omega}{2\pi i}\frac{d \beta}{2\pi i} \mathcal{D}A^{\alpha} \mathcal{D}A^{3} \, \mathcal{D}B\; e^{\ln\left(\beta\omega-\frac{\omega^{2}}{2}\right)} e^{\beta}\,e^{\frac{\omega}{2}} \exp \left[-\frac{1}{2} \int \frac{d^{4}p}{(2\pi)^{4}} A^{\alpha}_{\mu}(p)\, Q^{\alpha \beta}_{\mu \nu}\, A^{\beta}_{\nu}(-p)\right]  \nonumber \\
&\times&  \exp \left[-\frac{1}{2} \int \frac{d^{4}p}{(2\pi)^{4}} \left(
 \begin{array}{cc}
  A^{3}_{\mu}(p) & B_{\mu}(p)
 \end{array} \right)
\mathcal{P}_{\mu \nu} \left(
\begin{array}{ll}
A^{3}_{\nu}(-p) \\
B_{\nu}(-p)
\end{array} \right) \right]
\label{Zquad2}
\end{eqnarray}
where
\begin{eqnarray}
Q^{\alpha \beta}_{\mu \nu} &=& \left[ p^{2} + \frac{\nu^{2} g^{2}}{2} + \frac{g^{2}}{2}\beta \frac{1}{p^{2}} \right] \delta^{\alpha \beta} T_{\mu\nu}(p^2)\,,\quad \mathcal{P}_{\mu \nu} = \left(
                        \begin{array}{cc}
                         p^{2} + \frac{\nu^{2}}{2}g^{2} + \frac{\omega}{2}g^{2}\frac{1}{p^{2}} & - \frac{\nu^{2}}{2}g\,g' \\
                         - \frac{\nu^{2}}{2}g\,g' & p^{2} + \frac{\nu^{2}}{2}g'^{2}
                        \end{array} \right) T_{\mu\nu}(p^2)\;.
 \label{Qmunu}
\end{eqnarray}
One can easily get the two-point correlation function of the fields $A^{\alpha}$, $A^{3}$ and $B$ by inverting the matrices $\mathcal{Q}^{\alpha \beta}_{\mu\nu}$ and $\mathcal{P}_{\mu\nu}$. Thus, since $\mathcal{Q}^{\alpha \beta}_{\mu\nu}$ is diagonal, the off-dagonal boson propagator can be written as
\begin{equation}
\label{diagprop}
\langle A^{\alpha}_{\mu}(p)A^{\beta}_{\nu}(-p)\rangle = \frac{p^{2}}{p^{4} + \frac{\nu^{2}}{2}g^{2}p^{2} + \frac{\beta}{2}g^{2}} \delta^{\alpha\beta}T_{\mu\nu}(p^2)\;.
\end{equation}
Similarly, inverting $\mathcal{P}_{\mu\nu}$ immediately gives the diagonal boson propagators, namely
\begin{subequations} \label{aandbprops} \begin{gather}
\label{offdiagprop1}
\langle A^{3}_{\mu}(p)A^{3}_{\nu}(-p)\rangle = \frac{p^{2}(p^{2} + \frac{\nu^{2}}{2}g'^{2})}{p^{6}+\frac{\nu^{2}}{2}p^{4}(g^{2}+g'^{2})+\frac{\omega^{2}}{4}g^{2}(2p^{2}+\nu^{2}g'^{2})} T_{\mu\nu}(p^2)\;, \\
\label{offdiagprop2}
\langle B_{\mu}(p)B_{\nu}(-p)\rangle = \frac{p^{4} + \frac{\nu^{2}}{2}g^{2}p^{2} + \frac{\omega^{2}}{2}g^{2}}{p^{6}+\frac{\nu^{2}}{2}p^{4}(g^{2}+g'^{2})+\frac{\omega^{2}}{4}g^{2}(2p^{2}+\nu^{2}g'^{2})} T_{\mu\nu}(p^2)\;, \\
\label{offdiagprop3}
\langle A^{3}_{\mu}(p)B_{\nu}(-p)\rangle = \frac{\frac{\nu^{2}}{2}gg'p^{2}}{p^{6}+\frac{\nu^{2}}{2}p^{4}(g^{2}+g'^{2})+\frac{\omega^{2}}{4}g^{2}(2p^{2}+\nu^{2}g'^{2})} T_{\mu\nu}(p^2)\;.
\end{gather} \end{subequations}
With the relations \eqref{wsandza}, one obtains the propagators of $W^{\pm}$, $Z$ and $A$:
\begin{subequations} \label{alltheprops} \begin{gather}
\langle W^{+}_{\mu}(p)W^{-}_{\nu}(-p)\rangle = \frac{p^{2}}{p^{4} + \frac{\nu^{2}}{2}g^{2}p^{2} + \frac{\beta}{2}g^{2}} \delta^{\alpha\beta}T_{\mu\nu}(p^2)\;, \label{ww} \\
\langle Z_{\mu}(p)Z_{\nu}(-p)\rangle = \frac{p^{4} + \frac{\omega^{2}}{2}\frac{g^{2}g'^{2}}{(g^{2}+g'^{2})}}{p^{6}+\frac{\nu^{2}}{2}p^{4}(g^{2}+g'^{2})+\frac{\omega^{2}}{4}g^{2}(2p^{2}+\nu^{2}g'^{2})} T_{\mu\nu}(p^2)\;, \label{zz} \\
\langle A_{\mu}(p)A_{\nu}(-p)\rangle = \frac{p^{4} + \frac{\nu^{2}}{2}p^{2}(g^{2}+g'^{2}) + \frac{\omega^{2}}{2}\frac{g^{4}}{(g^{2}+g'^{2})}}{p^{6}+\frac{\nu^{2}}{2}p^{4}(g^{2}+g'^{2})+\frac{\omega^{2}}{4}g^{2}(2p^{2}+\nu^{2}g'^{2})} T_{\mu\nu}(p^2)\;, \label{aa} \\
\langle A_{\mu}(p)Z_{\nu}(-p)\rangle = \frac{\frac{\omega^{2}}{2}\frac{g^{3}g'}{(g^{2}+g'^{2})}}{p^{6}+\frac{\nu^{2}}{2}p^{4}(g^{2}+g'^{2})+\frac{\omega^{2}}{4}g^{2}(2p^{2}+\nu^{2}g'^{2})} T_{\mu\nu}(p^2)\;. \label{az}
\end{gather} \end{subequations}
In order to derive the gap equations, we first integrate out the fields to obtain
\begin{equation}
Z_\text{quad} = \mathcal{N} \int \frac{d \omega}{2\pi i}\frac{d \beta}{2\pi i}\,e^{\ln\left(\beta\omega -\frac{\omega^{2}}{2}\right)} e^{\frac{\omega}{2}}\,e^{\beta} \left[\det Q_{\mu \nu}^{ab} \right]^{-1/2}\, \left[ \det \mathcal{P}_{\mu \nu} \right]^{-1/2} \;. \label{Zquad3}
\end{equation}
The two determinants appearing can immediately be evaluated as
\begin{subequations} \begin{gather}
\ln\det P_{\mu\nu} = 3 \int \frac{d^{4}p}{(2\pi)^{4}} \log \left(p^4+\frac{\nu^2}2(g^2+g'^2)p^2+\frac\omega2g^2+\frac{g^2g'^2}4\frac{\nu^2\omega}{p^2}\right) \;, \\
\ln\det Q_{\mu\nu}^{ab} = 3\times2 \int \frac{d^{4}p}{(2\pi)^{4}} \log \left(p^{2} + \frac{\nu^{2}}{2}g^{2} + \frac{g^{2}}{2}\beta \frac{1}{p^{2}}\right) \;,
\end{gather} \end{subequations}
where we have used the fact that the elements of $P_{\mu\nu}$ commute with each other, allowing to compute the determinant in the usual way. The partition function can be written as
\begin{eqnarray}
Z_\text{quad} = \mathcal{N} \int \frac{d \omega}{2\pi i}\frac{d \beta}{2\pi i} e^{f(\omega, \beta)} \label{Zf eq}\,,\qquad f(\omega, \beta) = \frac{\omega}{2} + \beta 
- \frac12 \ln\det Q_{\mu\nu}^{ab} - \frac12 \ln\det P_{\mu\nu} \;.
\end{eqnarray}
Applying the saddle point approximation, one gets
\begin{equation}
Z_\text{quad} \simeq e^{f(\omega^{\ast}, \beta^{\ast})}\;,\quad \text{with}~ \frac{\partial f(\omega, \beta)}{\partial \beta}\bigg{|}_{\beta^{\ast}} = \frac{\partial f(\omega, \beta)}{\partial \omega}\bigg{|}_{\omega^{\ast}} = 0 \;.
\label{Zf eq1}
\end{equation}
For each condition we get a gap equations: the first one, form the $\omega$ derivative,
\begin{subequations} \label{gapeqs} \begin{equation}
\frac{3}{2}g^{2}\int\!\! \frac{d^{4}p}{(2\pi)^{4}}\,\frac{p^{2} + \frac{\nu^{2}}{2}g'^{2}}{p^{6} + \frac{\nu^{2}}{2}(g^{2} + g'^{2})p^{4} + \frac{\omega^{\ast}}{2}g^{2}p^{2} + \frac{\omega^{\ast}}{4}\nu^{2}g^{2}\,g'^{2}} = 1 \;,
\label{omega gap eq}
\end{equation}
and the second one, from the $\beta$ derivative,
\begin{equation}
\frac{3}{2}g^{2} \int \frac{d^{4}p}{(2\pi)^{4}} \frac{1}{p^{4}+\frac{\nu^{2}}{2}g^{2}p^{2}+\frac{g^{2}}{2}\beta^{\ast}} =1\;.
\label{beta gap eq}
\end{equation} \end{subequations}

\subsection{The limit $g' \to 0$.}
An important check to be done is the case where $g'=0$, which must recover the results of \cite{Capri:2012ah} with the Higgs field in the fundamental representation, obtaining a $U(1)$ massless gauge field decoupled from the $SU(2)$ gauge sector. This decoupling can be easily seen just by setting $g'=0$ in the propagator expressions \eqref{diagprop} and \eqref{aandbprops}:
\begin{subequations} \label{propgzro} \begin{eqnarray}
\langle A^{\alpha}_{\mu}(p)A^{\beta}_{\nu}(-p)\rangle &=& \frac{p^{2}}{p^{4} + \frac{\nu^{2}}{2}g^{2}p^{2} + \frac{\beta}{2}g^{2}} \delta^{\alpha\beta}T_{\mu\nu}(p^2)\;,\quad \langle A^{3}_{\mu}(p)A^{3}_{\nu}(-p)\rangle = \frac{p^{2}}{p^{4}+\frac{\nu^{2}}{2}g^{2}p^{2}+\frac{\omega^{2}}{2}g^{2}}T_{\mu\nu}(p^2)\;, \\
\langle B_{\mu}(p)B_{\nu}(-p)\rangle &=&  \frac{1}{p^{2}} T_{\mu\nu}(p^2)\;,\quad \langle A^{3}_{\mu}(p)B_{\nu}(-p)\rangle = 0\;.
\end{eqnarray} \end{subequations}
Also, as in the last section, one should be able to write the propagators in terms of the fields $W^{\pm}$, $Z$ and $A$ obtaining
\begin{subequations} \label{ppgzro} \begin{eqnarray}
\langle W^{+}_{\mu}(p)W^{-}_{\nu}(-p)\rangle &=& \frac{p^{2}}{p^{4} + \frac{\nu^{2}}{2}g^{2}p^{2} + \frac{\beta}{2}g^{2}} \delta^{\alpha\beta}T_{\mu\nu}(p^2))\;, \label{ww1} \quad \langle Z_{\mu}(p)Z_{\nu}(-p)\rangle = \frac{p^{2}}{p^{4}+\frac{\nu^{2}}{2}g^{2}p^{2}+\frac{\omega^{2}}{2}g^{2}} T_{\mu\nu}(p^2)\;, \\
\langle A_{\mu}(p)A_{\nu}(-p)\rangle &=& \frac{1}{p^{2}} T_{\mu\nu}(p^2)\;, \quad
\langle A_{\mu}(p)Z_{\nu}(-p)\rangle = 0\;.
\end{eqnarray} \end{subequations}
These propagators, \eqref{propgzro} and \eqref{ppgzro}, could also be derived by taking $g'=0$ in the quadratic partition function, or even in the generating function \eqref{Zquad1}, and following the steps of the last section.

\subsection{The vacuum energy}
Looking at the above propagators, beside the decoupling of the $U(1)$ gauge field from the $SU(2)$ gauge field, one should note the likeness between the diagonal and off-diagonal propagators, though in general the two Gribov parameters, $\omega^\ast$ and $\beta^\ast$, differ. Therefore, given the important role played by the gap equations, it seems to be worth to analyze the two gap equations \eqref{gapeqs} in the limit $g'\to 0$. In this limit, the gap equations \eqref{gapeqs} become
\begin{equation}
\label{omega gapeq g'zero}
\frac{3}{2}g^{2} \int\!\!\frac{d^{4} p}{(2\pi)^{4}}\, \frac{1}{p^{4}+\frac{\nu^{2}}{2}g^{2}p^{2}+\frac{\omega^{\ast}}{2}g^{2}} = 1\;, \qquad
\frac{3}{2}g^{2} \int\!\! \frac{d^{4}p}{(2\pi)^{4}}\, \frac{1}{p^{4}+\frac{\nu^{2}}{2}g^{2}p^{2}+\frac{g^{2}}{2}\beta^{\ast}} =1\;.
\end{equation}
It is clear that these two equations are identical. Thus, when $g' \to 0$, there is only one gap equation and, therefore, only one Gribov parameter. Consequently, the diagonal and off-diagonal propagators of \eqref{propgzro} are identical. These results coincide with what was found in \cite{Capri:2012ah} in the case of Higgs field in the fundamental representation.

Furthermore, from equation \eqref{Zf eq}, we get for the vacuum energy,
\begin{equation}
\mathcal{E}_{\omega^{\ast}} = \frac{3}{2}\omega^{\ast} - \frac{9}{2}g^2 \int\!\! \frac{d^{4}p}{(2\pi)^{4}}\log \! \left(p^{4} + p^{2}\frac{\nu^{2}}{2}g^{2} + \frac{\omega^{\ast}}{2}g^{2}\right)\;,
\end{equation}
which, again, is in agreement with the expression for the vacuum energy for the case of $SU(2)$ in the fundamental representation, upon redefining  $\frac{\omega^{\ast}}{2} = \frac{\vartheta^{\ast}}{3}$.

\section{About $\sigma_\text{off}(0)$ and $\sigma_\text{diag}(0)$ without the Gribov parameters} \label{sect3}
Before trying to solve the gap equations, it seems to be worthwhile to study what happens with $\sigma_\text{off}(0)$ and $\sigma_\text{diag}(0)$ in the absence of the Gribov parameters, which will allow us to search for regions where the Gribov parameters $\omega$ and $\beta$ are unnecessary, which happens whenever $\sigma_\text{off}(0)$ and/or $\sigma_\text{diag}(0)$ are less than one.

Thus, given \eqref{sigmaoffanddiag} and the propagators \eqref{AB gluon prop} we have
\begin{equation}
\label{sgoff} \langle \sigma_\text{off}(0) \rangle = \frac{3g^{2}}{4}\int\!\!\frac{d^{4}p}{(2\pi)^{4}}\,\frac{1}{p^{2}}\left(\frac{1}{p^{2}+\frac{\nu^{2}}{2}g^{2}}+\frac{1}{p^{2}+\frac{\nu^{2}}{2}(g^{2}+g'^{2})}\right)\,,\qquad
\langle \sigma_\text{diag}(0) \rangle = \frac{3g^{2}}{2}\int\!\!\frac{d^{4}p}{(2\pi)^{4}}\,\frac{1}{p^{2}}\frac{1}{\left(p^{2}+\frac{\nu^{2}}{2}g^{2}\right)}\;.
\end{equation}
Using standard techniques, this gives
\begin{equation}
\label{constas} \langle \sigma_\text{off}(0) \rangle = \frac{1}{2}\left(1-\frac{3g^{2}}{32\pi^{2}}\ln(2a)\right) + \frac{1}{2}\left(1-\frac{3g^{2}}{32\pi^{2}}\ln(2 a')\right)\,, \qquad
\langle \sigma_\text{diag}(0) \rangle = 1-\frac{3g^{2}}{32\pi^{2}}\ln(2a)\;,
\end{equation}
where
\begin{equation}
\label{consta} a = \frac{\nu^{2}g^{2}}{4\bar{\mu}^{2}\,e^{1-\frac{32 \pi^{2}}{3g^{2}}}}\,, \qquad a' = \frac{\nu^{2}(g^{2}+g'^{2})}{4\bar{\mu}^{2}\,e^{1-\frac{32 \pi^{2}}{3g^{2}}}} = a \frac{g^{2}+g'^{2}}{g^{2}} = \frac{a}{\cos^{2}(\theta_{W})}\;,
\end{equation}
such that the off-diagonal ghost form factor can be rewritten as
\begin{equation}
\langle \sigma_\text{off}(0) \rangle = 1 - \frac{3g^{2}}{32\pi^{2}}\ln\frac{2a}{\cos(\theta_{W})}\;.
\label{sigmaoff4}
\end{equation}
where $\theta_{W}$ is the Weinberg angle. With the 2nd expression of \eqref{sgoff} and \eqref{sigmaoff4} we are able to identify three possible regions:
\begin{itemize}
\item Region I, where $\langle \sigma_\text{diag}(0)\rangle < 1$ and $\langle \sigma_\text{off}(0)\rangle < 1$, meaning $2a > 1$. In this case the Gribov parameters are both zero so that we have the massive $W^{\pm}$ and $Z$, and a massless photon, as in \eqref{WZAgluonprop1}, in what we can call the Higgs phase.
\item Region II, where $\langle \sigma_\text{diag}(0)\rangle > 1$ and $\langle \sigma_\text{off}(0)\rangle < 1$, or equivalently $\cos \theta_{W} < 2a < 1$. In this region we have $\omega = 0$ while $\beta \neq 0$, leading to a modified $W^{\pm}$ propagator, and a free photon and a massive $Z$ boson.
\item The remaining parts of parameter space, where $\langle \sigma_\text{diag}(0)\rangle > 1$ and $\langle \sigma_\text{off}(0)\rangle > 1$, or $0 < 2a < \cos\theta_{W}$. In this regime we have both $\beta \neq 0$ and $\omega \neq 0$, which modifies the $W^{\pm}$, $Z$ and photon propagators, as shown in equations \eqref{alltheprops}. Furthermore this region will fall apart in two separate regions III and IV due to different behavior of the propagators.
\end{itemize}

\section{The off-diagonal gauge bosons} \label{sect4}
Let us first look at the behavior of the off-diagonal bosons under the influence of the Gribov horizon. The propagator \eqref{ww} only contains the $\beta$ Gribov parameter, meaning $\omega$ does not need be considered here.

As found in the previous section, this $\beta$ is not necessary in the regime $a>1/2$, due to the ghost form factor $\langle\sigma_\text{diag}(0)\rangle$ always being smaller than one. In this case, the off-diagonal boson propagator is simply of the massive type:
\begin{equation}
	\langle W^{+}_{\mu}(p)W^{-}_{\nu}(-p) \rangle = \frac{1}{p^{2} + \frac{\nu^{2}}{2}g^{2}}T_{\mu\nu}(p^2) \;.
\end{equation}

In the case that $a<1/2$, the relevant ghost form factor is not automatically smaller than one anymore, and the Gribov parameter $\beta$ becomes necessary. The value of $\beta$ is given by the gap equations \eqref{beta gap eq}, which has exactly the same form as in the case without electromagnetic sector. Therefore the results will also be analogous. As the analysis is quite involved, we just quote the results here.

For $1/e<a<1/2$ the off-diagonal boson field has two real massive poles in its two-point function. One of these has a negative residue, however. This means we find one physical massive excitation, and one unphysical mode in this regime. When $a<1/e$ the two poles acquire a nonzero imaginary part and there are no poles with real mass-squared left. In this region the off-diagonal boson propagator is of Gribov type, and the $W$ boson is completely removed from the spectrum. More details can be found in \cite{Capri:2012ah}.

\begin{figure}\begin{center}
\parbox{.5\textwidth}{\includegraphics[width=.5\textwidth]{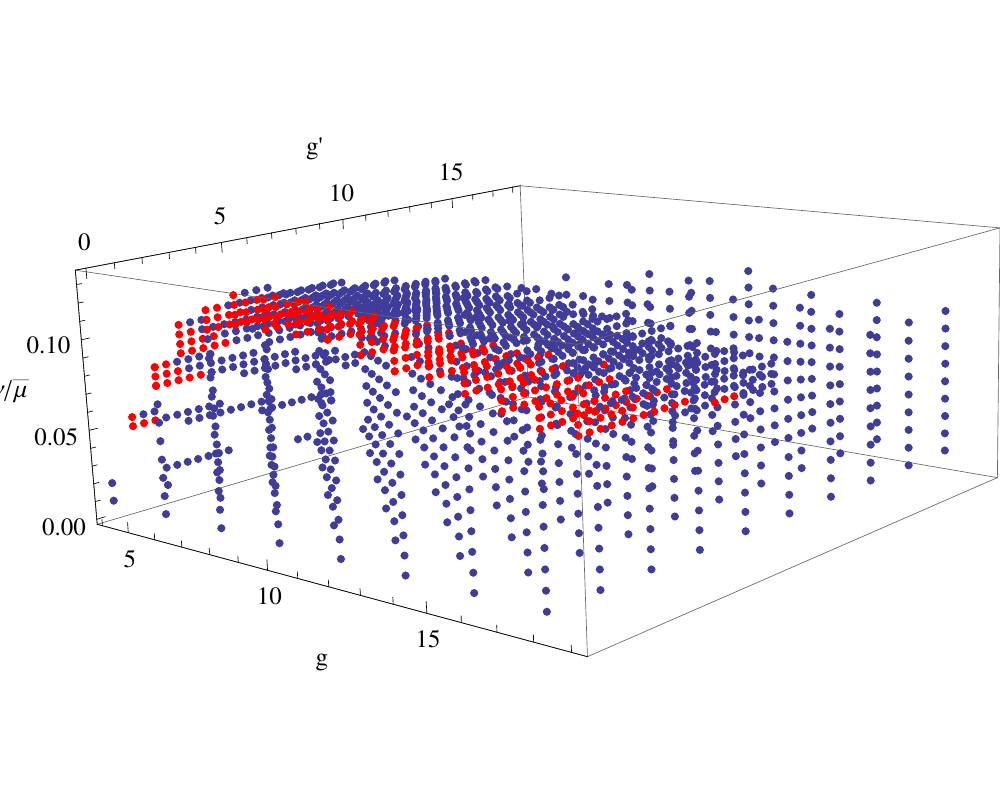}} \quad \parbox{.4\textwidth}{\includegraphics[width=.4\textwidth]{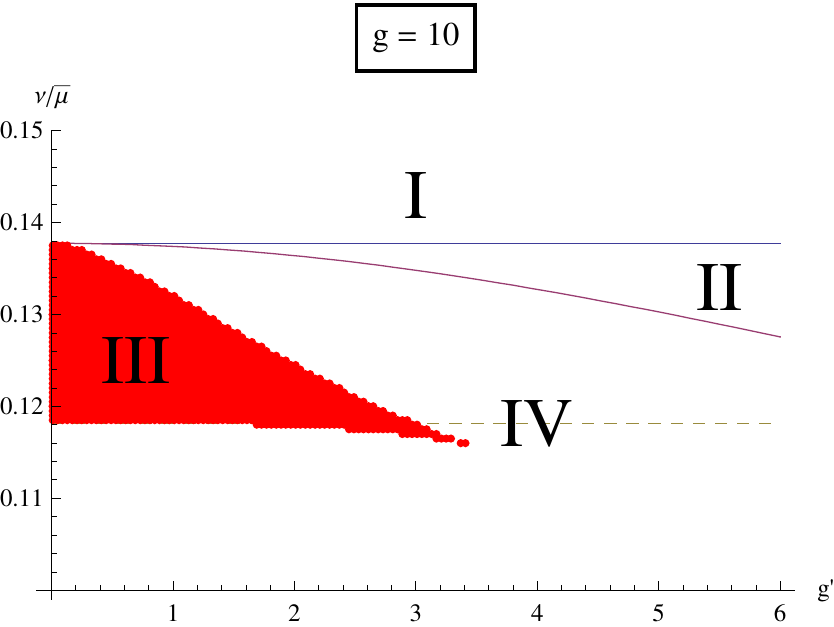}}
\caption{Left is a plot of the region $a'<1/2$ (the region $a'>1/2$ covers all points with higher $\nu$). In red are points where the polynomial in the denominator of \eqref{aandbprops} has three real roots, and in blue are the points where it has one real and two complex conjugate roots. At the right is a slice of the phase diagram for $g=10$. The region $a>1/2$ and $a'>1/2$ is labeled I, the region $a<1/2$ and $a'>1/2$ is II, and the region $a<1/2$ and $a'<1/2$ is split into the regions III (polynomial in the denominator of \eqref{aandbprops} has three real roots, red dots in the diagram at the left) and IV (one real and two complex conjugate roots, blue dots in the diagram at the left). The dashed line separates the different regimes for off-diagonal SU(2) bosons (two real massive poles above the line, two complex conjugate poles below). \label{regionsdiag}}
\end{center}\end{figure}

\section{The diagonal SU(2) boson and the photon} \label{zandgamma}
The two other gauge bosons --- the diagonal SU(2) boson and the photon --- have their propagators given by \eqref{offdiagprop1}, \eqref{offdiagprop2}, and \eqref{offdiagprop3} or equivalently --- the $Z_\mu$ and the $A_\mu$ --- by \eqref{zz}, \eqref{aa} and \eqref{az}. Here, $\omega$ is the only of the two Gribov parameters present.

In the regime $a'>1/2$, $\omega$ is not necessary to restrict the region of integration to $\Omega$. Due to this, the propagators are unmodified in comparison to the perturbative case:
\begin{equation} \label{propsrewrite}
	\langle Z_{\mu}(p) Z_{\nu}(-p) \rangle = \frac{1}{p^{2} + \frac{\nu^{2}}{2}(g^{2} + g'^{2})} T_{\mu\nu}(p^2)\;, \quad
	\langle A_{\mu}(p) A_{\nu}(-p) \rangle = \frac{1}{p^{2}}T_{\mu\nu}(p^2) \;.
\end{equation}

In the region $a'<1/2$ the Gribov parameter $\omega$ does become necessary, and it has to be computed by solving its gap equation. Due to its complexity it seems impossible to do so analytically. Therefore we turn to numerical methods. Using Mathematica the gap equation can be straighforwardly solved for a list of values for the couplings. In order to do this, we regularize the momentum integration by subtracting a term designed to cancel the large-$p^2$ divergence (as in the Pauli--Villars procedure):
\begin{equation}
	\frac{3}{2}g^{2}\int \frac{d^{4}p}{(2\pi)^{4}} \left(\frac{p^{2} + \frac{\nu^{2}}{2}g'^{2}}{p^{6} + \frac{\nu^{2}}{2}(g^{2} + g'^{2})p^{4} + \frac{\omega^{\ast}}{2}g^{2}p^{2} + \frac{\omega^{\ast}}{4}\nu^{2}g^{2}g'^{2}} - \frac1{(p^2+M^2)^2}\right) 	+ \frac{3}{2}g^{2}\int \frac{d^{4}p}{(2\pi)^{4}}\frac1{(p^2+M^2)^2} = 1 \;,
\end{equation}
where $M^2$ is an arbitrary mass scale. The second integral is readily computed by hand, whereas the first one converges and can be determined numerically. Once the $\omega$ parameter has been determined, we look at the propagators to investigate the nature of the spectrum.

The denominators of the propagators are a polynomial which is of third order in $p^2$. There are two cases: there is a small region in parameter space where the polynomial has three real roots, and for all other values of the couplings there are one real and two complex conjugate roots. In \figurename\ \ref{regionsdiag} these zones are labeled III and IV respectively.

Just as in \cite{threedcase} we can decompose the propagator matrix $\Delta_{ij}$ (were $i$ and $j$ run over the fields $A_\mu^3$ and $B_\mu$) as
\begin{eqnarray}
	\Delta_{ij} &=& \frac{f_{11}(-m_1^2)+f_{22}(-m_1^2)}{(-m_1^2+m_2^2)(-m_1^2+m_3^2)} \frac1{p^2+m_1^2} \hat v_i^1\hat v_j^1 \nonumber \\
	&+& \frac{f_{11}(-m_2^2)+f_{22}(-m_2^2)}{(m_1^2-m_2^2)(-m_2^2+m_3^2)} \frac1{p^2+m_2^2} \hat v_i^2\hat v_j^2 + \frac{f_{11}(-m_3^2)+f_{22}(-m_3^2)}{(m_1^2-m_3^2)(m_2^2-m_3^2)} \frac1{p^2+m_3^2} \hat v_i^3\hat v_j^3 \;. \label{partfracdec}
\end{eqnarray}
The vectors $v_i^n$ can be interpreted as linear combinations of the $A_3$ and $B$ fields. Decomposing the two-point functions in this way, we thus find three ``states'' $v_1^n A_3 + v_2^n B$. These states are not orthogonal to each other (which would be impossible for three vectors in two dimensions). The coefficients in front of the Yukawa propagators will be the residues of the poles, and they have to be positive for a pole to correspond to a physical excitation. From the analysis in \cite{threedcase} it also follows that $f_{11}(-m_n^2)+f_{22}(-m_n^2)$ will always be positive if the mass squared is real, which helps to determine the sign of the residue without having to explicitly compute it. The same results can also be attained by introducing ``generalized $i$-particles'', as done in \cite{threedcase} (see also \cite{Baulieu:2009ha}).

As the model under consideration depends on three dimensionless parameters ($g$, $g'$ and $\nu/\bar\mu$), it is not possible to plot the parameter dependence of these masses in a visually comprehensible way. Therefore we limit ourselves to discussing the behavior we observed.

In region III, when there are three real poles in the full two-point function, it turns out that only the two of the three roots we identified have a positive residue and can correspond to physical states, being the one with highest and the one with lowest mass squared. The third one, the root of intermediate value, has negative residue and thus belongs to some negative-norm state, which cannot be physical. All three states have nonzero mass for nonzero values of the electromagnetic coupling $g'$, with the lightest of the states becoming massless in the limit $g'\to0$. In this limit we recover the behavior found in this regime in the pure $SU(2)$ case \cite{Capri:2012ah} (the $Z$-boson field having one physical and one negative-norm pole in the propagator) with a massless boson decoupled from the non-Abelian sector.

In region IV there is only one state with real mass squared --- the other two having complex mass squared, conjugate to each other --- and from the partial fraction decomposition follows that it has positive residue. This means that, in this region, the diagonal-plus-photon sector contains one physical massive state (becoming massless in the limit $g'\to0$), and two states that can, at best, be interpreted as confined.

\section{Summary and Outlook} \label{concs}
We have taken a first look at the potential consequences of restricting the path integral gauge field configurations to the Gribov region, in order to deal with the inherent ambiguity in non-Abelian gauge fixing. We considered here the $4d$ $SU(2)\times U(1)$ theory.

The propagator of the gauge fields can get considerably modified by the presence of the nonperturbative Gribov parameters, in addition to the influence of the vacuum expectation value of the Higgs field. From the gauge boson pole structure, we distinguished a few regions in the $(g, \nu)$ parameter space for the physical spectrum of the theory, see in particular \figurename\ \ref{regionsdiag}.

For those values of the couplings and Higgs vacuum expection value which we expect perturbation theory to work for, we do recover the perturbative Yang--Mills--Higgs propagators. This is an important observation, as it means that the Gribov ambiguity does not spoil the physical vector boson interpretation of the gauge sector of the electroweak model at energies where it is relevant (i.e.~experimentally probed). For higher values of the non-Abelian coupling constant the Gribov horizon lets its influence be felt and the propagators become deeply modified, as was anticipated by the authors of \cite{Lenz:2000zt}. For nonzero electromagnetic charge, the $W$ bosons have a different behavior from the photon and $Z$ boson. For all gauge bosons, however, there is an intermediate region where a massive, physical pole still appears in the propagator, accompanied by a nonphysical pole with negative residue. At strong coupling all physical poles but one disappear (become complex) and it seems that only one massive particle is left in the spectrum. This massive state becomes the massless photon in the limit $g'\to0$, an important limiting case. As we predict here and in previous works \cite{Capri:2012cr,Capri:2012ah,threedcase} a quite rich ``propagator spectrum'' in terms of the parameters $(g,\nu)$, we hope this will stimulate further numerical studies of the propagators in such models \cite{Maas:2010nc}.

In the current work, the focus was on the spectrum of gauge bosons using the generalization of the semi-classical Gribov no-pole analysis. Topics we did not touch upon yet are the generalization to all orders of the restriction to the Gribov region (e.g.~following \cite{Vandersickel:2012tz} or \cite{Capri:2012wx}), the dynamical self-correction of the Gribov--Zwanziger action \cite{Dudal:2008sp} which affects the pole location amonst other things, and most importantly, the phase diagram of gauge--Higgs systems in terms of a variable interaction constant $g$ and Higgs expectation value $\nu$. This would allow an analytical continuum analysis of to what extent the renowned lattice predictions of Fradkin--Shenker \cite{Fradkin:1978dv,Greensite:2004ke,Caudy:2007sf,Greensite:2011zz} can be probed using the nonperturbative effective action approach that takes into account the Gribov ambiguity. The observation made here that depending on $(g,\nu)$, the behaviour of the propagators is considerably different although different regimes seem to be connected smoothly, suggests that there might be a connection to be unraveled w.r.t.~the analyticity predictions of \cite{Fradkin:1978dv}. Recent work in the Schwinger--Dyson formalism is in nice agreement with this observation \cite{Schaden:2013ffa}. However, in order to establish such a connection, the study of a suitable (quasi)order parameter like the Polyakov loop \cite{Greensite:2011zz,Fukushima:2012qa} would be needed\footnote{For the subtlety of the connection between the gauge propagators and the physical spectrum/phase diagram, see for example \cite{frohlich,BanksFrohlichMaas}}.

\section*{Acknowledgments}
The Conselho Nacional de Desenvolvimento Cient\'{\i}fico e
Tecnol\'{o}gico (CNPq-Brazil), the Faperj, Funda{\c{c}}{\~{a}}o de
Amparo {\`{a}} Pesquisa do Estado do Rio de Janeiro, the Latin
American Center for Physics (CLAF), the SR2-UERJ,  the
Coordena{\c{c}}{\~{a}}o de Aperfei{\c{c}}oamento de Pessoal de
N{\'{\i}}vel Superior (CAPES)  are gratefully acknowledged. D.~D.~is supported by the Research-Foundation Flanders.

\end{document}